\documentclass[
 amsmath,amssymb,
 reprint,%
]{revtex4-1}

\usepackage{geometry}
\usepackage{graphicx}
\usepackage{dcolumn}
\usepackage{bm}
\usepackage{textcomp}
\usepackage{appendix}

\begin{document}


\title[Digital Beings as an option to study gut flora evolution and adaptation]{Digital Beings as an option to study gut flora evolution and adaptation}

\author{Jo\~ao Paulo A. de Mendon\c ca}
 \email{jpamendonca@fisica.ufjf.br}
\author{Leonardo M. V. Teixeira}
\author{Fernando Sato}%
\affiliation{ 
Departamento de F\' isica \\
Universidade Federal de Juiz de Fora, Minas Gerais, Brazil}%

\date{\today}

\begin{abstract}
In this work, we introduce a computational model for the study of the host-bacteria interaction and the influence of the intestinal microbiota on the behavior and feeding pattern of an individual. The model is based on digital entities, which we've called Digital Beings (DBs), modeled using dynamic systems and genetic algorithms. We have successfully tested the use of the DBs by reproducing observation in previously made studies using rats and humans. Among these studies, we highlight those on how the bacteria in an individual's stomach could influence their eating behavior and how a controlled and continuous diet can affect the longevity of a certain population. Our results point that the Digital Beings can be used as a tool for supporting the devising of experiments and corroborating with theoretical hypotheses, reducing the number of \textit{in vivo} tests.
\end{abstract}

\maketitle

\section{\label{sec:intro}Introduction}

Food is strongly related to a given peoples' history, society and religion\cite{kiple2000cambridge,tannahill1988food}, being a main feature in the definition of cultural identity\cite{kittler2011food, montanari2006food}. Thinking of feeding simply as one of the mechanism that sustains our metabolism is a very simplified and inaccurate model. Nowadays, with global growth of food related problems such as obesity\cite{wang2007obesity,puhl2009stigma,swinburn2011global}, food intolerance\cite{nwaru2014prevalence} and starvation\cite{graham1993starvation}, our relation to what we eat deserves a special attention of the scientific community. The understanding of this relation can be the key for a more sustainable and healthier future for humanity.

There is a popular conception that says: the food that we eat define us. Recent studies, however, have pointed to an extension of this conception by showing the important rule that gut flora represents in food craving and intolerance\cite{alcock2014eating,rezzi2007human, de2013intestinal}. Also, it has already been shown that there is a relationship between gut flora and human mood or behavior\cite{schroeder2007antidepressant,bercik2011intestinal}. Some studies also pointed that the systemic ingestion of a kind of food by a person can cause specific bacteria to develop as part of its gut flora, as it's the case of seaweed\cite{hehemann2010transfer} or other cellulose rich foods\cite{de2010impact}. The scientific community is now looking at food with a new perspective, focusing on the host-microbiota relationship mediated by it.

Comprehending this relationship can be very advantageous for the future of medicine, since a research front points towards a gut flora based therapy for obesity\cite{castaner2018gut, kallus2012intestinal}, diabetes\cite{wen2008innate, cani2008changes}, bowel syndrome\cite{tana2010altered}, liver diseases\cite{le2012intestinal,schnabl2014interactions,mouzaki2013intestinal} and even depression\cite{lotrich2011role, bailey2011exposure}, among many others that have already been tested. These therapies are usually done by means of the use of probiotics, prebiotics and antibiotics in order to modulate the population of a specific bacteria in the gut flora.

The majority of those studies are made in live specimens, usually rats, and then subsequently tested in humans. Others choose to work with \textit{in vitro} tests before the \textit{in vivo} ones. This is expected, since there is little theoretical support for the hypotheses and explanations in this field. Part of this can be related to the fact that it is a new, thus under development, area of research. The systems focused are very complex and show many interactions that are already being discovered and debated by the  scientific community.   

Now, computer science have been used as an important ally in understanding medicine and biology problems or systems\cite{nasica2015amyloid,hallet2015systematic,portell2018microscale}. Simulations have already been used to model bacterial development and complex interactions in ecologic systems, with highlight to studies using dynamical systems\cite{gonze2018microbial,bucci2016mdsine}, game theory\cite{lambert2014bacteria} and evolutionary algorithms\cite{heilbuth2015estimation}.

In this paper we propose a theoretical model that captures the main features of the actual proposed mechanisms of host-gut bacteria interactions. Our model combines two main features: The first is a dynamical system to simulate the bacteria-host relation inside a hypothetical host that we have called a Digital Being (DB). The second is an evolutionary algorithm that models the changes in a DB population which appear due to external pressure to which it is exposed. Our model was tested by comparing to previous results obtained from experiments on rats on previously published studies, regarding alimentary compulsion and lifespan. Our results show that the DB can be used efficiently as a platform to test hypotheses in this kind of study, giving theoretical feedback, reducing the overall costs and cutting down on the number of living animal subjects.

\section{\label{sec:method}Methodology}

The proposed model that we studied is based on some basic assumptions on the nature of the host-microbiota system, which aims to represent the physical reality of the system:

\begin{enumerate}
\item \textbf{Three bacteria populations} compose the DB (host) gut flora. The number of specimens for each kind of bacteria at a given moment $t$ shall be called $A(t)$, $B(t)$ and $C(t)$.  
\item The DB diet is composed basically of \textbf{three types of food}, each of which favors individually one of the three kinds of bacteria. The proportion of the type of food associated to bacteria $A$ that the DB ingests at a given moment $t$ will be called $N_A(t)$. The same notation shall be used for the other two types of food.
\item \textbf{The host eats at regular intervals}, such that we associate to each individual DB two characteristic times: $t_{fr}$ (time interval at which the bacteria environment is food rich) and $t_{fp}$ (time interval at which the bacteria environment is food poor).
\item \textbf{The bacteria populations decay over time naturally}, with a half-life (the time needed for a given bacteria population to become halved) which we shall call $t^A_{\text{\textonehalf}}$ for population $A$. In the dynamical system, we shall make use of a set of constants, namely $\alpha_{A}$ for population A, representing the decay rate of the bacteria population. The same notation shall be used for populations $B$ and $C$.
\item \textbf{When exposed to a food rich environment, the populations of bacteria grow over time}, with a doubling-time (the time needed for a given population to become twice what it was) we shall call $T^A_{2}$ for population $A$. The growth rate is influenced linearly by the abundance of the respective bacteria's given type of food (our $N$ values), with the proportionality constant given by the $\beta$'s in the dynamical system. The same notation shall be used for populations $B$ and $C$.
\item There is a mechanism through which the \textbf{bacteria can influence the DB eating behavior}. This process can be treated in many ways, and we shall use a simple set of constants ($\gamma_A$, $\gamma_B$ and $\gamma_C$) to regulate this mechanism. These constants also incorporate the interaction with any DB related agent other than the three bacteria we are working with, e.g. other bacteria populations, or even hypothetical diseases.
\item \textbf{Extinction or overpopulation of any type of bacteria will kill the host DB}. We call extinction when a population fall under 0.1\% of the initial average population (i.e., $(A(0)+B(0)+C(0))/3$) and overpopulation when a population gets bigger than a value 1000 times this initial average.  
\end{enumerate}

When considering a different system, different assumptions may have to be made. It is important to understand that these assumptions are in fact what we consider as hypotheses. If the simulations don't behave exactly like the system that's being simulated, it's probably because one or more of the considerations we made are, in fact, wrong or that we forgot something important. This is one of the best payoffs given by numerical models: A wrong hypothesis usually means loss of implementation and processing time, coupled with the human effort made in putting the ideas into the model. If this hypothesis test had been done via an experiment, a wrong model could represent a larger cost, which usually comprehends human working time summed with laboratory materials and, sometimes, animal lives.

\begin{figure}
    \centering
    \includegraphics[width=7cm]{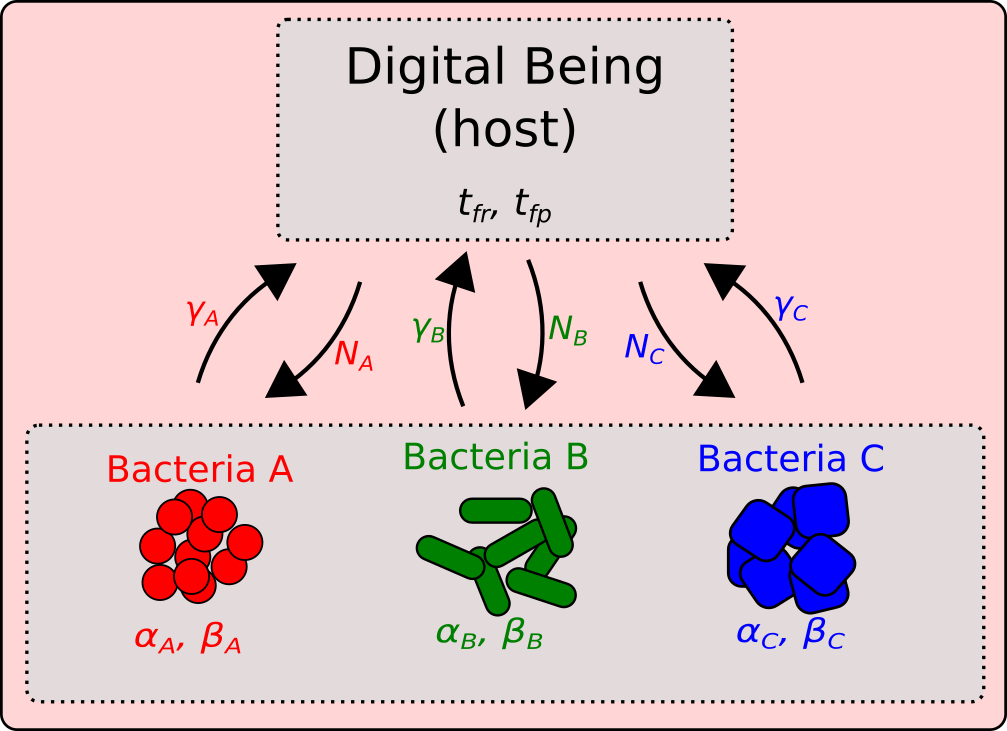}
    \caption{Schematic representing the interactions we introduced in the dynamical system that models a DB's life-cycle. Even if they are being represented separately, it is important to realize that all these variables are in fact emergent from the adaptation between host and bacteria, as can be seen in section \ref{sec:appl}. At the same time, we can say the $\alpha$'s and $\beta$'s characterize a given bacteria population in the same manner that $t_{fr}$ and $t_{fp}$ characterizes the host DB.}
    \label{fig:scheme}
\end{figure}

A graphic and schematic representation of all the constants we introduced in the model can be seen in figure \ref{fig:scheme}. In modeling the life of one individual DB and its interactions with their gut flora, we first need to define the equations that represent the features presented as the assumptions. In order to make our set of time evolution equations easier, we calculate the following constant associated to our proposed values:

$$
\alpha_A = \frac{\log(2)}{t^A_{\text{\textonehalf}}}
$$

$$
\beta_A = \frac{\log(2)}{T^A_{2}}
$$

In which $\log$ denotes the natural logarithm. The same equations are valid for populations $B$ and $C$. The mechanism of food craving that responds to bacteria stimulus will be modeled by a simple proportional division, where the weight is the bacteria stimulus strength, which we are representing as our $\gamma$ constants, multiplied by the bacteria population at the moment the meal starts. The food preferences of the DB is directly associated to the amount of food available to the bacteria in its gut flora, so it can be better represented by this variable. So, the food availability for bacteria will be modeled as the following example for population $A$ related food:

$$
N_A (t) = 
\begin{cases} 
\frac{\gamma_A A(t_0)}{\gamma_A A(t_0)+\gamma_B B(t_0)+\gamma_C C(t_0)}, & \mbox{if with food} \\ 
0, & \mbox{if without food} 
\end{cases}
$$

Where $A(t_0)$, $B(t_0)$ and $C(t_0)$ express the $A$, $B$ and $C$ populations at the moment in which the respective DB meal will start. The same equations apply to populations $B$ and $C$. Considering these variables previously defined, and our assumptions about the system studied, we can write the following set of time evolution equations for the DB bacteria populations, properly written for Euler integration:

$$
\begin{cases}
    A(t+dt) = A(t)[1 + (\beta_A N_A (t) - \alpha_A)dt]\\
    B(t+dt) = B(t)[1 + (\beta_B N_B (t) - \alpha_B)dt]\\
    C(t+dt) = C(t)[1 + (\beta_C N_C (t) - \alpha_C)dt]\\
\end{cases}
$$

At a first glance these equations seem independent of each other, but in fact we have six dynamic equations: Three for population evolution and three for food availability evolution. From this, one can derive a set of differential equations for the same problem, but we shall work with the equations as shown, which are more suitable for a direct numerical implementation. As convention, we choose as standard values $dt=0.01{\text atu}$ and $t_{\text max}=24000.00{\text atu}$, where ${\text atu}$ stands for \textit{arbitrary time unit}, which we shall  omit from now on. 

For direct applications, the values have to be chosen in real world units. For example, one can express $A$ in CFU/mL and $t^A_{\text{\textonehalf}}$ in seconds. Anyway, we will focus on the behavior of the DBs, not directly comparing the values with real world ones. That being said, we will present our variables as being dimensionless. 

\section{\label{sec:appl}Application and results}

\begin{figure}
    \centering
    \includegraphics[width=7cm]{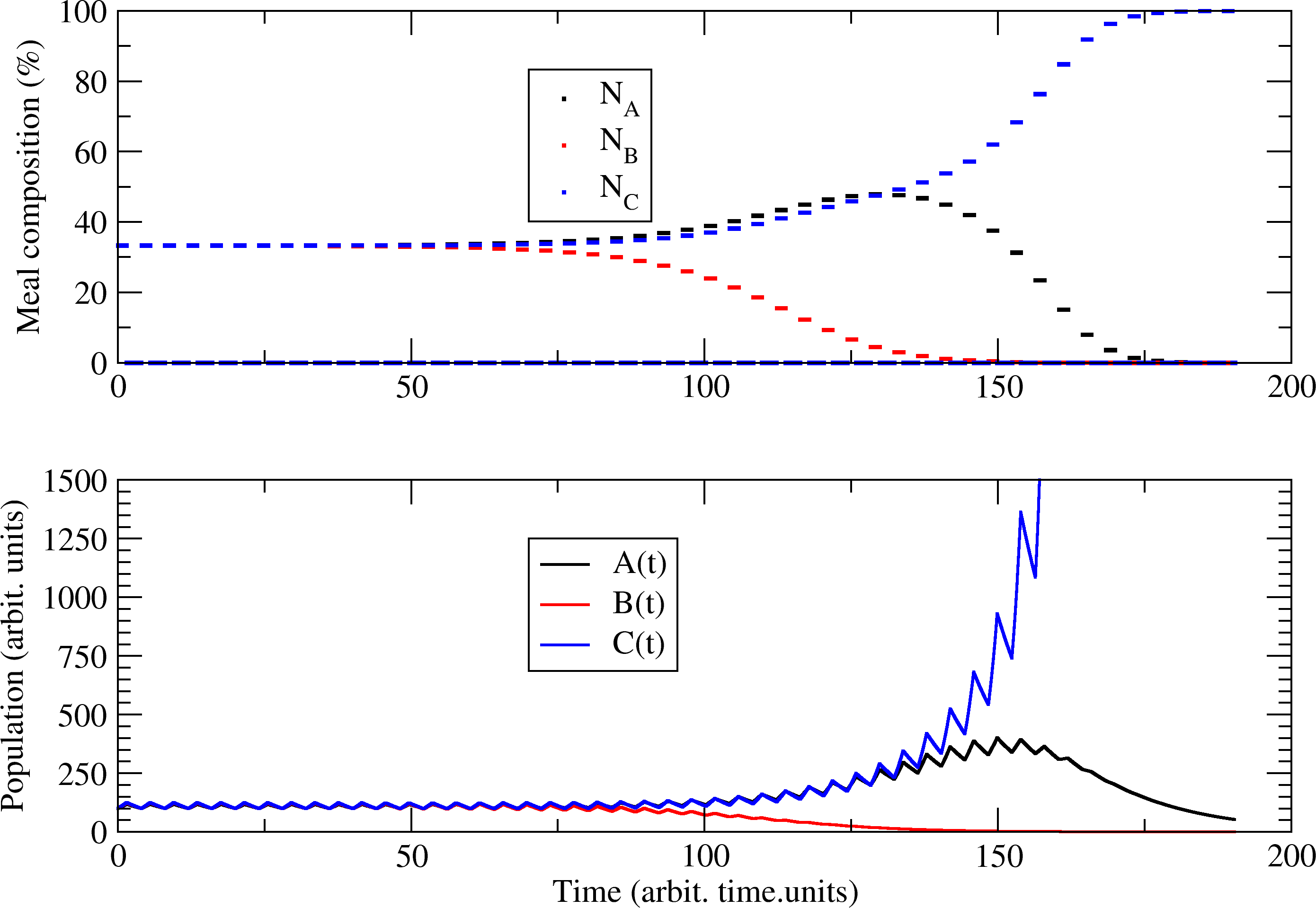}
    \caption{Profile of the functions N and of the bacteria populations in a life cycle of a DB that suffers from a excessive growth of C bacteria population, which generates a total suppression of B bacteria favorite food ingestion. The parameters used are $A_0=100$, $B_0=100$, $C_0=100$, $t^A_{\text{\textonehalf}}=10.4$, $t^B_{\text{\textonehalf}}=8.0$, $t^C_{\text{\textonehalf}}=7.6$, $T^A_{2}=1.3$, $T^B_{2}=1.0$, $T^C_{2}=0.95$, $\gamma_A=220.0$, $\gamma_B=220.1$ and  $\gamma_C=220.2$. The feeding pattern is set to consist of an interval of $1.5$ of full followed by $2.5$ of empty gut.}
    \label{fig:doenca}
\end{figure}

Let's now apply the model to a test problem. First, we will show the evolution of the bacteria populations in the life cycle of a DB. For this, we choose arbitrary values for the $\alpha$'s, $\gamma$'s and $\beta$'s, as well as the DB initial bacteria population and feeding cycle. The set of values and the graphical behavior of the dynamical variables can be seen in fig. \ref{fig:doenca}.

As the values were set such that bacteria C influences a little more in food preferences, the other species were rapidly suppressed by C, even with their larger population variation rates. It elucidates the already proposed idea that food preferences play a positive feedback cycle with bacteria population. However, as $\gamma$ is a variable that derives from the host-bacteria relationship, it is natural to think that evolutionary processes can change this value in order to enlarge the lifespan of our DB.

To test the previous hypothesis, we set an evolutionary algorithm that starts with a random 1000 DBs population and evolves it, positively selecting the DBs that have survived the longer. Through 1000 generations, the 25\% DBs with longer lifespans were chosen to generate asexually 4 offspring each. Normally, one feature of the DB son differs from its parent in as much as 2\%, preserving all the other features identical. So, with a chance of 0.2\%, we considered a total mutation of this random characteristic, assuming completely independent values from the ones in its DB parent. $A$, $B$ and $C$ were set to be in a range from 10.0 to 50.0 in the first step of each generation. Also,  $0.5<t_{\text{\textonehalf}}<50.0$, $0.5<T_{2}<50.0$, $100.0<\gamma<1000.0$, $t_{fr}+t_{fp}=5.0$ and $0.5<t_{fp}<4.5$.

\begin{figure}
    \centering
    \includegraphics[width=7cm]{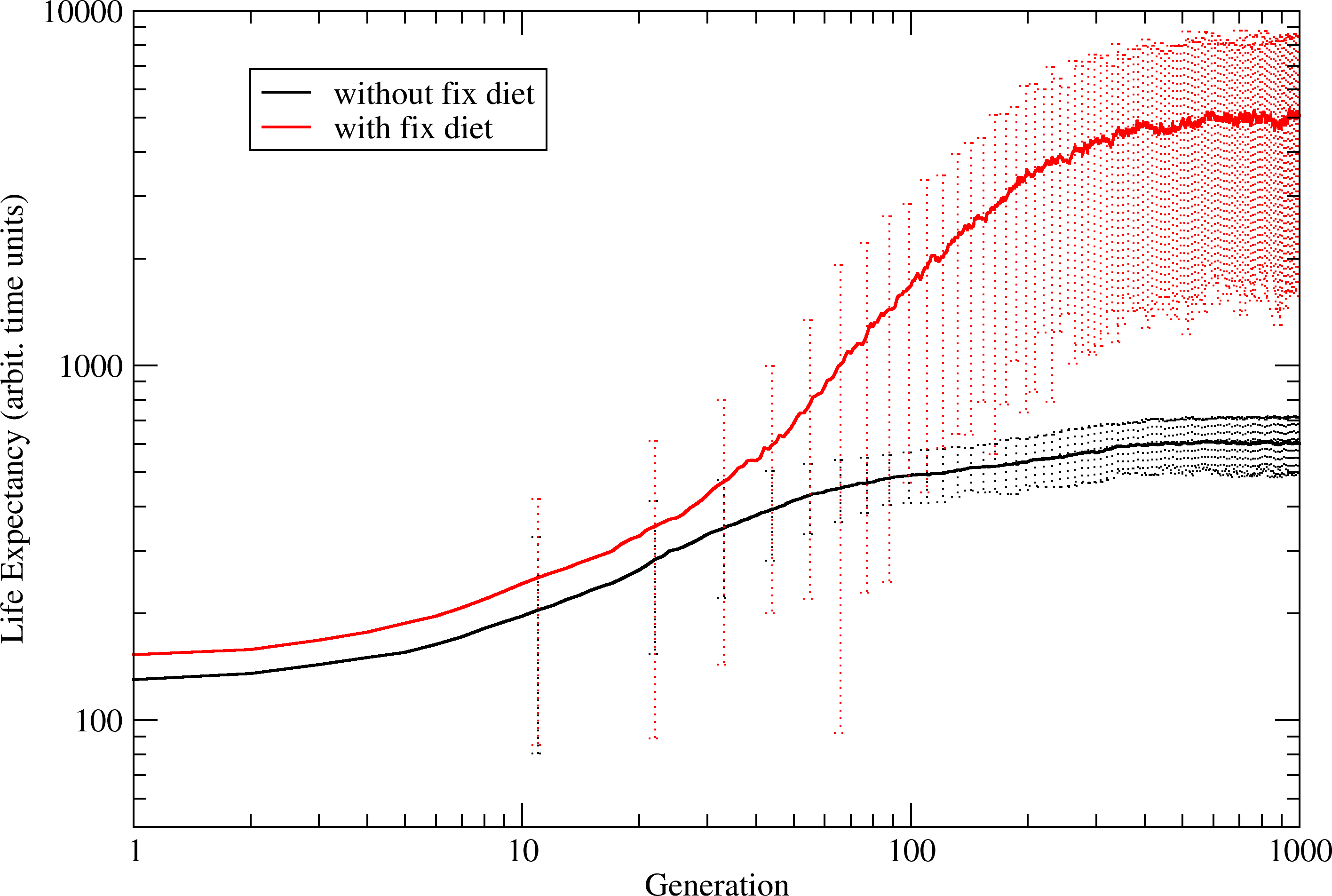}
    \caption{Lifespan behavior of DB populations with controlled and uncontrolled diets, with standard deviation, during the evolutionary process.}
    \label{fig:lifespan}
\end{figure}

The behavior of this DB population's lifespan over the generations can be seen as the black line in fig. \ref{fig:lifespan}. The results successfully pointed that DB populations can evolve over time and adapt their relationship with bacteria. After 1000 generations, the evolution was really slow, and the population assumed a lifespan of $602.2\pm105.6$.

Other possible hypothesis that can be tested at this point is that populations of individuals that are less susceptible to diet fluctuations, following a solid and constant diet over life, can evolve into larger lifespans than the one we observed previously. To test this, we repeated the evolutionary process described before, excluding the $\gamma$'s and making $N_i(t)=N_{0i}$ for $i=A,B,C$. The $N_{0i}$ values are let to evolve, being always larger than 0.1 and preserving the relation $N_{0A}+N_{0B}+N_{0C}=1.0$. 

As can be seen again in fig. \ref{fig:lifespan}, the same behavior we saw before was repeated, but this time with an immediate gain in life span, which grows over the generations, stabilizing in a lifespan of $5061.1\pm3498.7$. Our results point that DB populations that eat following controlled diets can evolve longer lifespan over generations than DBs from a population that follows an impulsive diet (in this case, generated by bacteria). 

On the other hand, it revealed more than that: We observed that long lifespans seems to be a characteristic associated to fixed diets, which is only intensified in the evolutionary process. This can be exemplified when looking to the first random populations, which have a lifespan of $145.8\pm123.6$ eating a rigid diet, while $124.0\pm100.4$ eat a bacteria induced one. So, even confirming our hypothesis, we observed that the evolutionary process simply intensifies the effects that are already present when we compare both random initial populations. 

\section{\label{sec:conclusion}Conclusions}

Attending to the crescent attention that the host-microbiota relationship is gaining in the literature, our aim here was to give support to this important field of research. In this work, we presented a theoretical and computational tool to test hypotheses and study problems that arrive from the long and complex interactions between the gut flora and its host's brain. We developed a model based on the dynamical systems' description of a hypothetical specimen, which we called a Digital Being, and its relation with its gut flora. From the tests, we observed that the DBs indeed exhibit the food craving caused by host-bacteria interaction, and its susceptibility in the choice of the parameters $\gamma$.

Combining this model with an evolutionary algorithm, we were able to study the impact of this relationship through generations and see how the mutual adaptations develop. We performed tests over two controlled populations of DBs: One evolved with microbiota induced eating behaviors and other with a fixed diet followed over all its life. From these tests, we confirmed that the population with a fixed diet evolved to an approximately ten times larger lifespan than the population that is subject only to bacteria induced behavior. This confirmed our hypothesis that the evolutionary process can enhance the variables of the specimens in order to maximize lifespan, but also pointed that fixed diets usually induce longer lives in random populations, independent of evolution itself. The evolution only intensifies this difference. All these results are in good agreement and give new insights into previous works on the literature\cite{alcock2014eating}. As our aim in this work was just to show the efficiency of our methodology, we shall not delve further into this matter.

With this in mind, we are presenting our DB based model for bacteria-host interaction to the scientific community. Since DB simulations are very adaptable and can be written and run on average computers, we hope that it can become an useful tool to support researchers in biology and medicine, by giving them theoretical support and controlled conditions' tests over proposed hypothesis, reducing costs and saving animal lives. 

We, the authors, acknowledge to Capes, Fapemig, CNPq, Finep and UFJF for all the support provided on this work. We are also tankful to our colleagues from the Departamento de F\'isica - UFJF, for the time spent in discussions that indirectly helped on this work.

\bibliographystyle{plain}
\bibliography{ref.bib}

\newpage
\begin{appendices}
\section*{Appendix: Detailed Analysis}

\subsection{Initial Populations}

\begin{figure*}[htb]
    \centering
    \includegraphics[width=15cm]{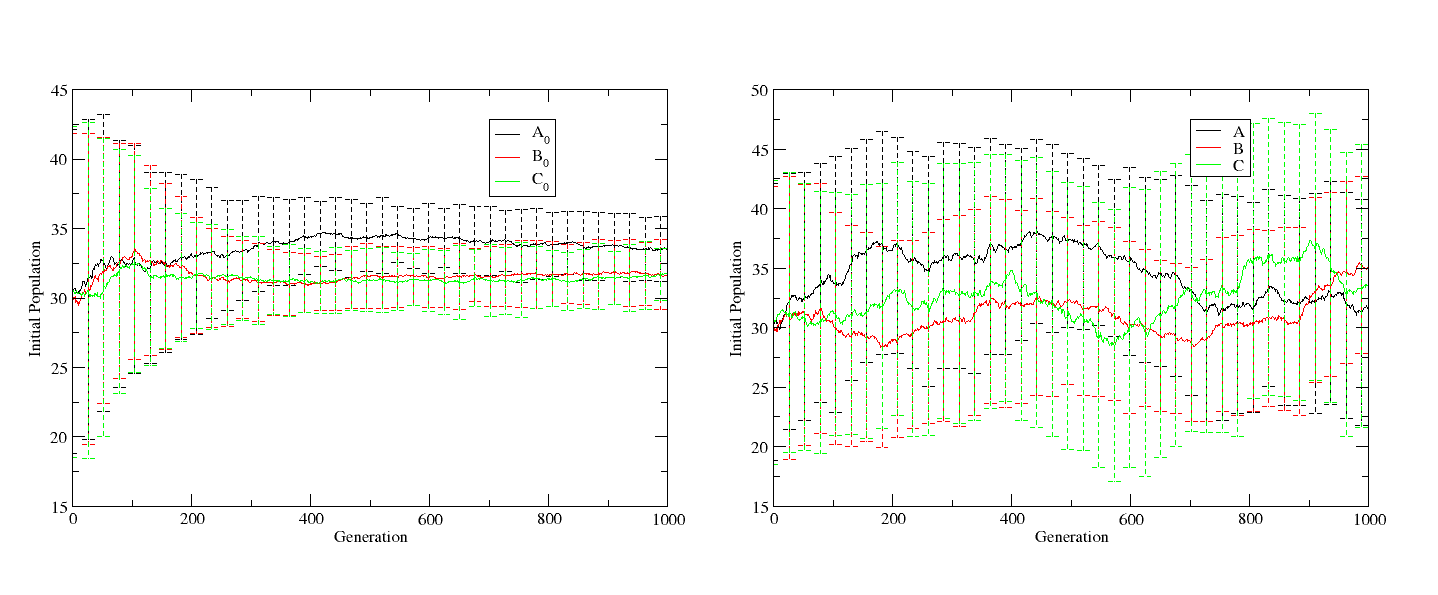}
    \caption{Evolution of the initial populations for the standard DB (left) and the diet DB (right). When taking the standard deviation into consideration, the difference in the values doesn't prove too significant.}
    \label{fig:inipop}
\end{figure*}

We begin our analysis with the \textit{initial population} parameters. The graphs showing the evolution of their mean values, for both DB types, can be seen in Figure \ref{fig:inipop}. From the diet DB graph, we see that even though the mean values change slightly over time, with no very significant changes observed, but the deviation are such that, in the end, it can be considered that no difference in the initial populations occur. For the standard DB, however, one of the bacteria attains a greater initial population, while the population of the other two evolve together with the same value. 
This difference in the evolution behavior may be related to a longer lifespan, but given that for the diet DBs, the ones with the longest lifespan, the change in the initial population is not significant, the influence should be minimal.

\subsection{Half-life Time}

\begin{figure*}[htb]
    \centering
    \includegraphics[width=15cm]{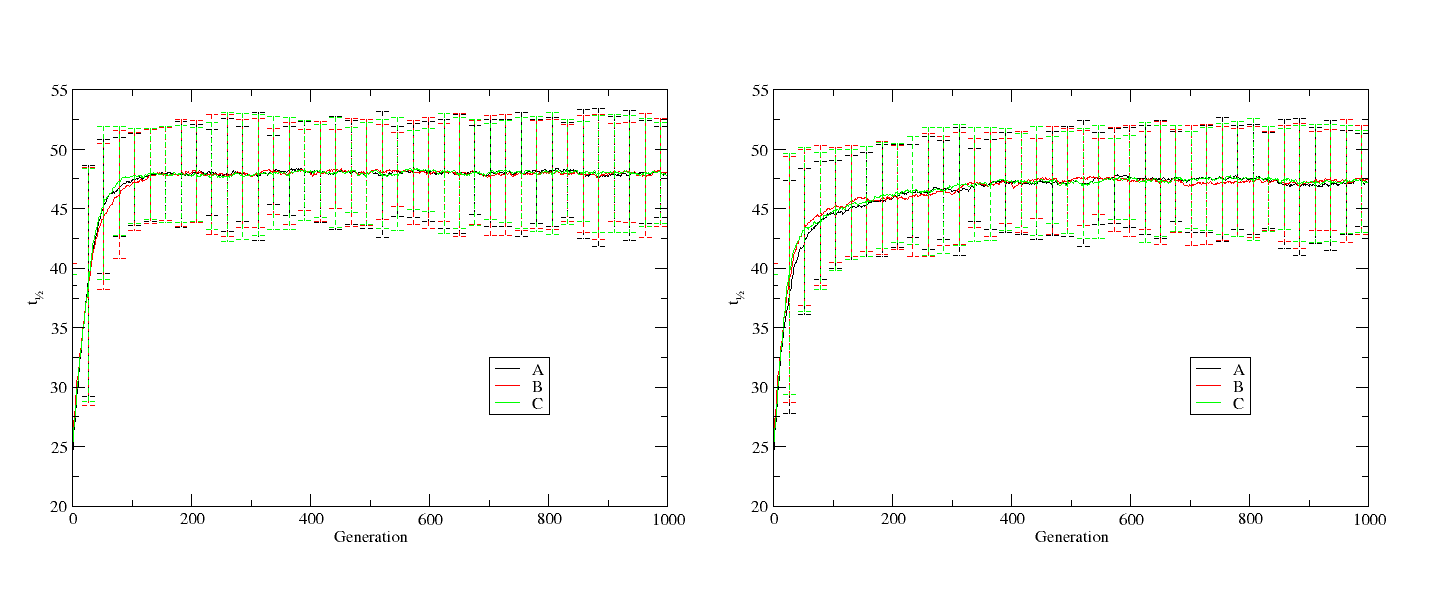}
    \caption{Graphs showing the evolution of the half-time parameter for the standard DB on the left and for the diet DB on the right. Their overall behavior is the most striking characteristic that can be extracted.}
    \label{fig:halftime}
\end{figure*}

The next parameter is the the \textit{half-life time}. This parameter is associated with the reduction of the bacteria populations throughout the dynamics, and thus pose an important parameter in the equilibrium of the gut flora in the Digital Beings. The graphs displaying the evolution can be seen in Figure \ref{fig:halftime}. 

The behavior observed in the evolution of this parameter is analogous for both types of DB. There is a significant increase that takes place still at the initial generations, in which the value of the parameter almost saturates the upper limit stipulated for the simulation, and is kept balanced throughout the rest of the evolution. This behavior happens for all three kinds of bacteria, and the value attained is also virtually the same. It indicates that this parameter is essential to the increase in the lifespan of the DBs, and higher values are most valued.

There is, however, a small difference between the standard and diet DB parameters. The increase in the value for the standard DB happens faster, whereas for the diet DB the parameter takes a longer time to reach it's final value. This hints that the increase in the parameter for the diet DB happens in greater balance with changes in other parameters, and in the standard DB it is more independent.

\subsection{Doubling Time}

\begin{figure*}[htb]
    \centering
    \includegraphics[width=15cm]{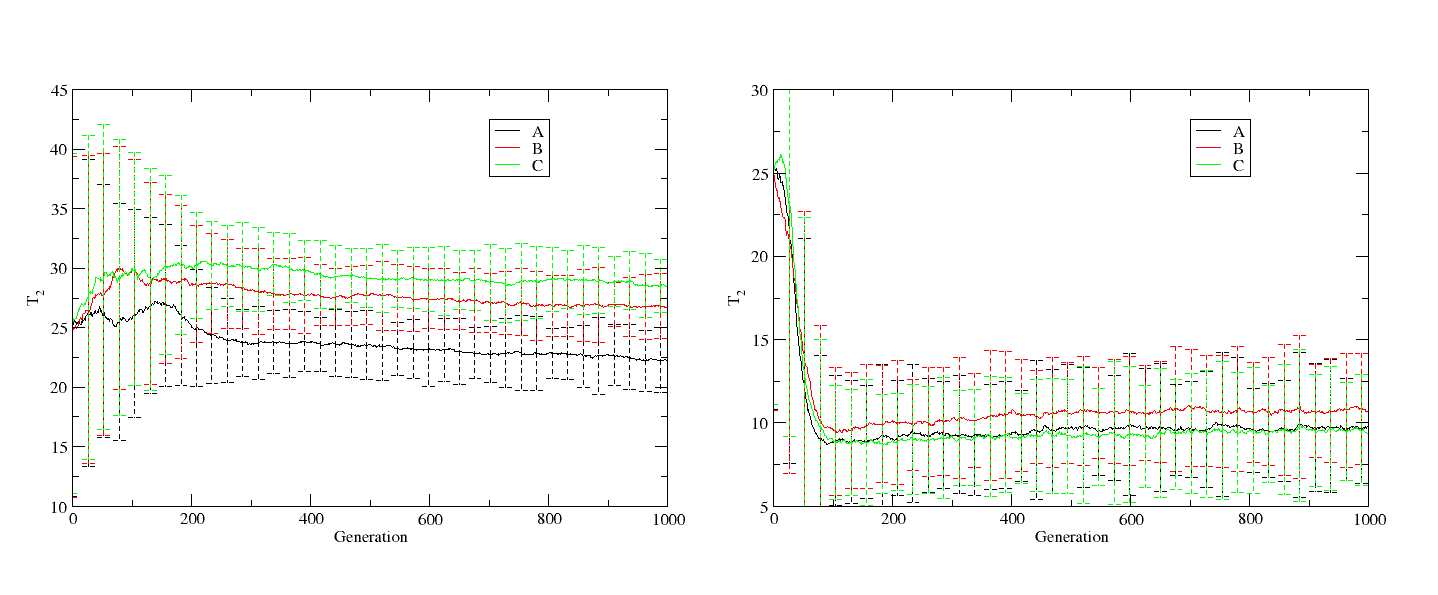}
    \caption{Evolution of the doubling time parameter for the standard DB (left) and the diet DB (right). Notice the difference in the pattern on the graphs.}
    \label{fig:doubtime}
\end{figure*}

After looking at the half-life time, the question on the behavior of the doubling time naturally arrives. The \textit{doubling time} dictates the growth rate of the bacteria in the gut flora, and together with the half-life time, help to maintain a population balance in the DBs bacteria numbers, being an important factor in the lifespan, remembering that the causes of death of the DBs are either overpopulation or disappearance of a given bacteria in its microbiota. The graphs presenting its evolution are shown in Figure \ref{fig:doubtime}.
Unlike the half-life time, the behavior of the doubling time is completely different for each DB type. 

The most striking detail is that the behavior of the doubling time, for the diet DBs, is similar to that of the half-life time, but now with a sudden reduction in the first generations, with a later slow evolution to its final value. This final value, however, is far from the lower limit imposed on the simulation, which is $0.5$. Notice also that, albeit the mean value for one of the populations differ, when considering the standard deviation we see that they are substantially the same, and they evolve together for all populations.

For the standard DB, the difference is radical compared to the diet DB. The overall deviation in the values is significantly smaller, with two populations showing an overall increase in the value, with the same deviation compared to initial values. One of the populations achieve a different value -- smaller -- than the other two, even considering the deviation. Notice also that the values do not stabilize at a value, like the diet DB, but slowly change until the end of the evolution. The mean values form a pattern that shall be explored further in the conclusions. 

\subsection{Full and Empty Stomach Time}

\begin{figure*}[htb]
    \centering
    \includegraphics[width=15cm]{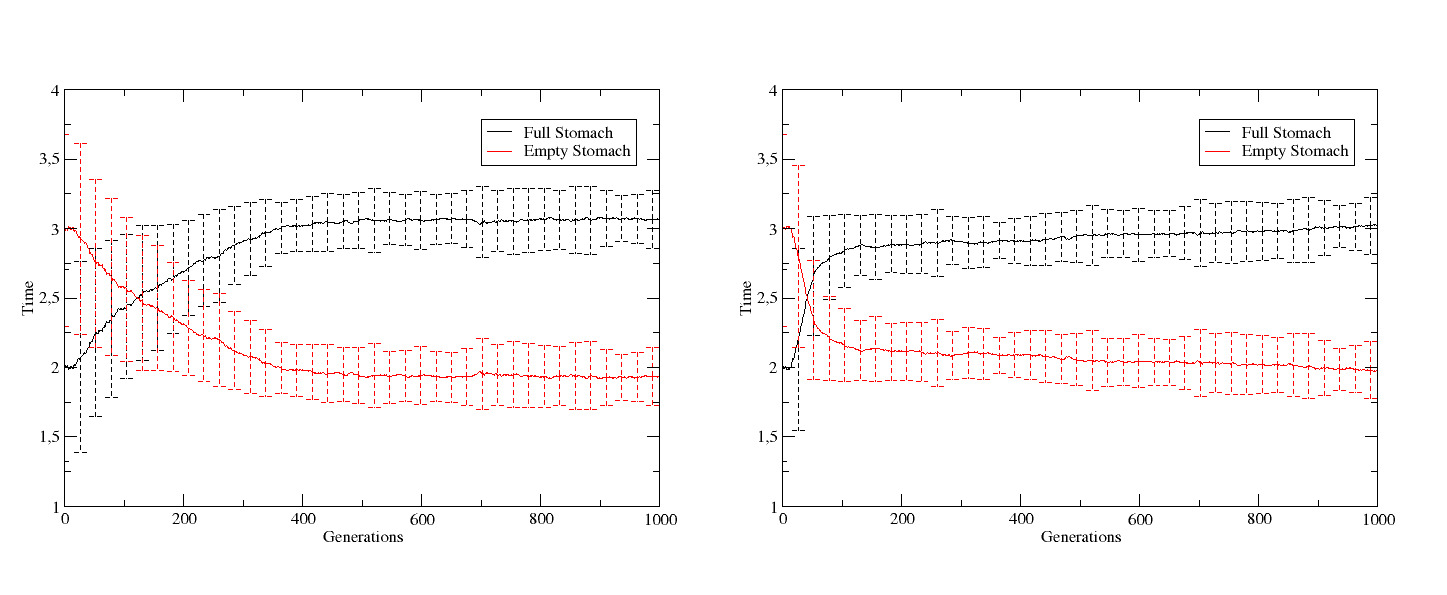}
    \caption{Evolution of full and empty stomach times. In the left, we see the graph showing the evolution for the standard DB. In the right, the graph showing the evolution for the diet DB.}
    \label{fig:stomtime}
\end{figure*}

The doubling time determines the rate in which the number of bacteria increase, but they only proliferate when there is food on the DB's stomach. Therefore, the amount of time in which the DB has, or not, food in its stomach is important. This characteristic is regulated by the \textit{full stomach time} and \textit{empty stomach time} parameters, and as such it is important to understand their evolution. The graphs showing their evolution are shown in Figure \ref{fig:stomtime}.

The same overall behavior is seen for both DB types: the empty and full stomach time values are reversed, remembering that the values are constrained by their sum. Nonetheless, the details of the evolution is different for each type of DB. For the diet DBs, the reversal happens more suddenly, but the subsequent change happens slowly, with the total reversal -- the moment when one parameter assumes the initial value of the other -- happening only at later generations. The value of the parameters also do not stabilize: there is a non-zero inclination in both curves. 

For the standard DBs, the reversal is slower, taking a longer time to happen, but the total reversal of the parameters happens much faster than that of the diet DBs. After the total reversal, those parameters increase until a final value, at which they remain fixed throughout the rest of the evolution. Notice that again we see a pattern in which the diet DB's parameters increase slower than the standard DB parameters. We shall delve further into this detail in the conclusions. 

\subsection{Feeding Pattern}

\begin{figure*}[htb]
    \centering
    \includegraphics[width=15cm]{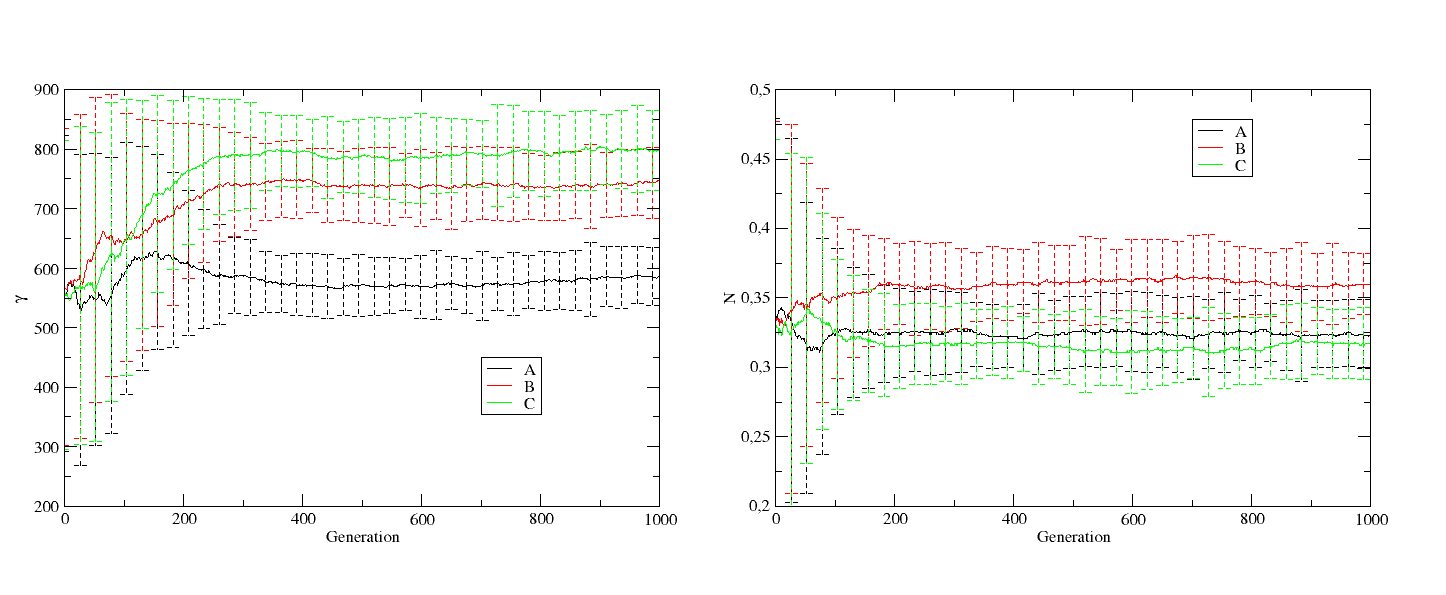}
    \caption{Graphs depicting the evolution of the feeding pattern for the standard DB (left) and for the diet DB (right).}
    \label{fig:diet}
\end{figure*}

At last, the analysis is not complete without talking about the \textit{feeding pattern}. It is the most important parameter, around which the entire study revolves, important to both the growth and decline of the bacteria populations, and the one that most deeply regulates it, without being a characteristic dependent on the bacteria alone. We now look at its evolution in dept.

The feeding pattern is governed by different parameters in the standard and diet DBs. In the former, the proportion of each kind of food that the DBs ingest throughout a generation varies during its entire life, and cannot be used as a representative of the generation's feeding pattern. Nonetheless, these are influenced by the $\gamma$ parameters, which represent the sensibility of the DB to the tastes of the bacteria. These parameters are dependent on the bacteria in their gut microbiota, and thus are fixed throughout an entire DB generation, therefore being the best parameter to represent the evolution of the feeding pattern for the standard DBs.

The diet DBs, as their name suggest, follow a strict diet, therefore their sensibility to the bacteria preferences is irrelevant (indeed, the $\gamma$ parameters are zero for the diet DBs). Nonetheless, since their diet is fixed for each DB in a given generation, this diet itself represent their feeding pattern. In terms of the parameters used in the simulation, we are talking about the proportion of each kind of food that they ingest. 

This difference in the parameters representing the feeding pattern is the main difference in the characterization of the DB types, and as mentioned before, is central to the study. The graphs showing the evolution of the feeding pattern, each type of DB with its relevant parameter, can be seen in Figure \ref{fig:diet}.

For the diet DBs, they develop a preference for the kind of food associated to one type of bacteria, represented by the greater amount of that kind of food ingested by the DBs. As for the other two, the difference is two slim that do not overcome the error bars, meaning that there is no significant change in interest between one another. The final pattern is a gap between two of the food preferences and the third one.

For the standard DBs, the scenario is different. We observe that the DBs develop a greater taste for the kind of food associated with two distinct bacteria populations, with this taste competing, as the difference in their values is no greater than their deviation. The taste for the food associated with the third bacteria is significantly smaller than the others, meaning it is less desired by the DBs. This is the opposite to the scenario seen for the diet DBs, in which we observe only one dominant bacteria, with the other two competing at smaller values. 

This completely opposite result is one of a kind among the parameters, and is very important. Remembering that after the evolution the diet DBs can achieve lifespan values almost 10 times the greater values achieved by the standard DB, the difference shown in their feeding behavior poses the only factor with a significant enough difference to justify this result. None of the other parameters have shown a difference so drastic to justify the obtained results.

\section{Conclusions}
Throughout the text we analyzed the evolution of all the main parameters of our models, in order to compare their behavior for both the standard and the diet Digital Beings, and to obtain their relation to the increase in lifespan for the DBs. First we note that none of the bacteria had any special advantage over the others, thus what is relevant in the observed results aren't the numbers or the characteristic for every bacteria, but the overall pattern and behavior shown by the parameters.

From the analysis, we observed that the \textit{initial population} doesn't seem to have a strong influence in he increase in lifespan, given the fact that its variation was slight, and in the diet DB scenario no significant variation was seen whatsoever. In order to have a long lifespan, the bacteria in the DB's gut must take a long time to die, as evidenced by the spike in their \textit{half-life time} values, and no bacteria must take longer to die than the other. Besides, they can't take too long to multiply, as shown by the evolution of the \textit{doubling time} parameter: the shorter the time they take to multiply, the longer the DBs are expected to live, as evidenced in the diet DB parameter. 

Coupled to that, it is clear that the DBs must spend most of their time with a \textit{ful stomach}, and a lesser time with an \textit{empty stomach}, in order to allow for the bacteria population to stabilize. Most importantly, however, its the \textit{feeding pattern}, which governs all of the above. There was a total opposition on the patterns, and from the diet DB pattern there's a hint that, for a longer lifespan, only one bacteria preference should be allowed.

These results were drawn by considering the parameters one by one. The biggest, and most interesting results, however, are obtained when we look at the evolution of the parameters as a whole, and compare the outcomes. By doing hat, we see that not only a longer half-life time and a smaller doubling time contributes to a longer lifespan, but a combination of both represent the optimal condition for the longest lifespans. On top of that, but the doubling time pattern is directly associated to that of the feeding pattern: the bacteria who receives the most amount of food is the one that takes longer to multiply, and vice-versa. This can be seen clearly when comparing both the pattern in the doubling time graphs with those on the feeding pattern graphs. One more feature that we can observe is that a longer adaptation time among the parameters seem to be key in allowing the hosts to have a longer lifespan. This longer adaptation time can be seen in the \textit{half-life time}, \textit{doubling time} and \textit{full and empty stomach time} parameters for the diet DBs, whereas the change in the standard DB generations is faster. These characteristics combined form the structure for a long-lived Digital Being.

Nonetheless, it is important to note that these results are only valid for the single simulation we presented in the text. The genetic algorithm rely on probabilities and chance, this means that it is not deterministic. Take, for instance, the fact that even thought no bacteria had any advantage over the others, they presented different behaviors. This means that if we were to do another run of the simulation, we could expect completely different results as outputs. Thus, for the study we made, the results presented above are relevant, but if we were trying to define the reasons and characteristics that undermine the long-living characteristic of a population, we wouldn't have enough data. If one wished to do such a study, an statistical study comparing many evolution trajectories and extracting their key characteristics would prove extremely significant, granting more general results and information, and providing a deeper insight on the mechanisms governing the interaction between the lifespan and the gut microbiota of a living being.
\end{appendices}

\end{document}